\documentclass[oldversion]{aa}
\usepackage{natbib}
\usepackage{graphicx}
\usepackage{txfonts}

\begin{document}

\title{Signatures of restarted activity in core-dominated,\\
       triple radio sources selected from the FIRST survey}

\author{A. Marecki\inst{1}
   \and P. Thomasson\inst{2}
   \and K.-H. Mack\inst{3}
   \and M. Kunert-Bajraszewska\inst{1}
   }
\titlerunning{Signatures of restarted activity in FIRST radio sources}
\offprints{Andrzej Marecki \email{amr@astro.uni.torun.pl}}
\institute{Toru\'n Centre for Astronomy, N. Copernicus University,
           87-100 Toru\'n, Poland
    \and   Jodrell Bank Observatory, The University of Manchester,
           Macclesfield, Cheshire, SK11 9DL, UK
    \and   INAF -- Istituto di Radioastronomia, Via Gobetti 101,
           40129 Bologna, Italy}

\date{Received 5 August 2005 / Accepted 21 October 2005}

\abstract{Signatures of the re-occurrence of activity in radio-loud AGNs,
indicated either by the so-called double-double or X-shaped structures,
have been observed in a number of radio sources. All such objects known to
date have linear sizes of the order of a megaparsec. A number of the sources
that are appreciably more compact than this, but that exhibit hints of a
past phase of activity, were found in the VLA FIRST survey. Their structures
show symmetric relic lobes straddling relatively bright, unresolved cores.
Observations of the cores of 15 such structures with MERLIN at
5\,GHz have shown that four of them are doubles or core-jets on the
subarcsecond scale. Misalignments of $\Delta PA \ga 30\degr$ between the axis
of the inner structure and the line connecting the fitted maxima of the
arcminute-scale relic lobes are clearly visible in three of the four sources.
From these results, we can infer that a rapid repositioning of
the central engine in each of these three radio sources is the most
plausible interpretation of the observed morphology and that a merger is
most likely the original cause of such a repositioning. In the case of
TXS\,1033+026, the optical image extracted from the SDSS archives clearly
suggests that two objects separated by only 2.7\,kpc (projected onto the 
sky plane) are indeed merging. The inner parts of TXS\,0818+214 and 
TXS\,1312+563 could be interpreted as double-lobed, and consequently, these
sources could be of the double-double type; but further multifrequency
observations are necessary to provide support for such an interpretation.}


\maketitle


\section{Introduction}\label{s-intro}

Firm observational evidence now exists that a galaxy, regardless of its
state of activity, hosts a supermassive black hole \citep[SMBH, see
e.g.][and references therein]{barth03}. If a significant amount of matter in
the vicinity of the SMBH accretes onto it, the galaxy appears to be active,
but for how long does this activity occur? Can an inactive galaxy become
active and vice versa, and is repeated switching between active and inactive
states possible? Radio observations can perhaps give some answers to these
questions because the lobes in radio-loud AGNs partly provide a historical
record of past activity. From detailed measurements of the spectral
indices of their lobes, the spectral ages of extended radio sources can
be inferred that, according to \citet{al87} and \citet{liu92}, are of the
order of $\sim$$10^7$ to $10^8$~years. However, as discussed in detail by
\citet{scheuer95}, double-lobed radio sources are known to expand with
velocities about a few percent of the speed of light;
consequently the overall linear sizes of these sources must be of the order
of $10^5$ to $10^6$\,pc. This agrees with observations and the fact that
some radio galaxies -- they are termed Giant Radio Galaxies (GRGs) -- may
have sizes $\ge 1$\,Mpc\footnote{For consistency with numerous papers on
GRGs \citep[e.g.][]{schoen99, schoen00a, schoen00b, schoen00c, kaiser00,
lara01} $H_0$=50\,km\,s$^{-1}$\,Mpc$^{-1}$ and $q_0=0.5$ is used throughout
this paper. For these cosmological parameters the redshift value of $z=1.25$
provides the maximum linear size for a given angular size: $1\arcsec$
corresponds to 8.6\,kpc.}. As the spectral ages of radio galaxies, even GRGs,
are thought to be two orders of magnitude less than their actual ages, which, in
turn, can be about the same age as the Universe \citep{pasq04}, it
would appear that activity phenomena are episodic: they can be started at an
arbitrary stage in the evolution of the galaxy and can cease or become
interrupted after a certain period of time.

If the energy supply from the central engine to the hotspots and the lobes
of a radio-loud AGN cuts off, it enters a so-called ``coasting phase''. It
can remain at this stage of its evolution for up to $10^8$\,years, and its
spectrum would become very steep due to radiation and expansion losses during
this period \citep{kg94,sl01}. On the other hand, as first pointed out by
\citet{hr74}, the radiative lifetimes of the relativistic electrons in the
hotspots are of the order of only $10^4$\,years. Therefore, the lack of
fuelling in radio sources that have switched off -- they are sometimes
termed ``faders'' -- should result in the absence of
the hotspots as they should have faded away relatively quickly. A few
objects of this kind have been observed in samples of ultra-steep spectrum
sources \citep{rott94, breuck00} or as a part of very low frequency surveys
\citep{cohen04,mack05}.

\begin{table*}[t]
\begin{flushleft}
\caption[]{Sources observed using MERLIN at 5\,GHz.}
\begin{tabular}{l c c r r c l}
\hline
\hline
\multicolumn{1}{c} {Source} & R.A.& Dec.&
\multicolumn{2}{c} {Integrated flux at} & {Spectral} &
\multicolumn{1}{c} {Sub-arcsecond} \\
&
\multicolumn{2}{c} {(J2000)} &
\multicolumn{1}{c} {1.4 GHz} &
\multicolumn{1}{c} {4.85 GHz} & {index} &
\multicolumn{1}{c} {structure}\\
\multicolumn{1}{c} {(1)} & (2) & (3) &
\multicolumn{1}{c} {(4)} & (5) & (6) & \multicolumn{1}{c} {(7)} \\
\hline

TXS 0726+256		& 07 29 05.744 & +25 30 34.24 &  136.40 &  63 & 0.64 & single\\
TXS 0818+214		& 08 21 07.501 & +21 17 03.06 &  187.55 &  56 & 1.00 & Figure~\ref{fig:0818+214}\\
TXS 0940+001		& 09 43 19.237 & $-$00 04 25.08 & 1158.61 & 499 & 0.70 & single\\
TXS 1002+554		& 10 06 17.423 & +55 12 44.98 &  169.80 &  62 & 0.83 & single\\
TXS 1024+549		& 10 27 34.175 & +54 42 40.25 &  118.29 &  48 & 0.75 & undetected\\
TXS 1025+089		& 10 28 33.021 & +08 39 57.12 &  103.18 &  40 & 0.78 & undetected\\
TXS 1033+026		& 10 36 31.933 & +02 21 45.53 &  235.71 & 120 & 0.56 & Figure~\ref{fig:1033+026}\\
TXS 1046+187		& 10 48 53.954 & +18 29 37.65 &   90.68 &  46 & 0.56 & marginally detected\\
TXS 1308+011		& 13 11 21.170 & +00 53 19.23 &  257.97 &  98 & 0.80 & Figure~\ref{fig:1308+011}\\
TXS 1309+484		& 13 12 11.144 & +48 09 25.22 &  165.55 &  85 & 0.55 & single\\
TXS 1312+563		& 13 14 58.416 & +56 03 42.00 &  364.00 & 155 & 0.71 & Figure~\ref{fig:1312+563}\\
FIRST J152523.6+053736	& 15 25 23.639 & +05 37 35.94 &  233.34 & 113 & 0.60 & single\\
FIRST J155726.1+360133	& 15 57 26.130 & +36 01 33.24 &  136.27 &  45 & 0.92 & marginally detected\\
FIRST J162544.9+271929	& 16 25 44.872 & +27 12 29.39 &  293.30 & 110 & 0.81 & single\\
B3 1704+437		& 17 06 24.098 & +43 40 30.37 &  123.33 &  45 & 0.84 & single\\

\hline
\end{tabular}
\end{flushleft}
{\small
Description of the columns:
Col.~(1): Source name;
Col.~(2): Right ascension (J2000) extracted from FIRST;
Col.~(3): Declination (J2000) extracted from FIRST;
Col.~(4): The sum of the fluxes densities of FIRST components;
Col.~(5): GB6 flux density;
Col.~(6): Spectral index calculated using data from Cols. 4 and 5;
Col.~(7): Sub-arcsecond structure as imaged using MERLIN at 5\,GHz
and reported in this paper --\\see the figures and notes in
Subsection~\ref{s-comm}.
}
\label{t-sample}
\end{table*}

\begin{table*}[t]
\begin{flushleft}
\caption[]{Positions and 1.4\,GHz flux densities of the components
of four sources with nontrivial sub-arcsecond structures
as indicated by the FIRST catalogue.}

\begin{tabular}{c c c r r c}
\hline
\hline
Source & R.A. & Dec. &
\multicolumn{2}{c }{$S$~[mJy]} &
\multicolumn{1}{c}{Distance from} \\
{} &
\multicolumn{2}{c}{(J2000)} &
\multicolumn{1}{c}{Peak} &
\multicolumn{1}{c}{Total} &
\multicolumn{1}{c}{the `core' [\arcsec]} \\
\hline

		& 08 21 06.547  & +21 17 19.26  &    6.92   &    23.25  &   21.0 \\
TXS 0818+214	& 08 21 07.501  & +21 17 03.06  &  142.20   &   148.32  &    --- \\
		& 08 21 08.357  & +21 16 49.09  &    6.12   &    15.98  &   18.4 \\
		&               &               &           &           &       \\
		& 10 36 28.166  & +02 22 28.88  &    4.15   &     4.11  &   71.2 \\
TXS 1033+026	& 10 36 31.933  & +02 21 45.53  &  186.11   &   202.43  &    --- \\
		& 10 36 33.024  & +02 21 36.80  &   21.32   &    27.92  &   18.5 \\
		&               &               &           &           &       \\
		& 13 11 20.744  & +00 53 10.70  &   28.22   &    51.67  &   10.7 \\
TXS 1308+011	& 13 11 21.170  & +00 53 19.23  &  134.20   &   144.57  &    --- \\
		& 13 11 21.786  & +00 53 29.46  &   39.40   &    61.73  &   13.8 \\
		&               &               &           &           &       \\
		& 13 14 57.908  & +56 03 32.46  &   18.54   &    28.08  &   10.4 \\
		& 13 14 57.987  & +56 03 46.06  &    2.39   &    20.69  &    5.4 \\
TXS 1312+563 	& 13 14 58.416  & +56 03 42.00  &  270.24   &   276.44  &    --- \\
		& 13 14 59.232  & +56 03 52.20  &   23.95   &    34.39  &   12.3 \\

\hline
\end{tabular}
\end{flushleft}
\label{t-FIRST-comps}
\end{table*}

Although it is obvious that total disappearance should normally be the ultimate
stage in the evolution of a fader, it is also clear that under certain
circumstances a new period of activity could begin during the ``coasting
phase'', i.e. before the lobes have faded out completely. An observable effect
of this would be the presence of new, bright components located in or straddling
the centre of a larger, double-lobed relic structure. The signature of such
renewed activity is most convincing if it takes the form of a smaller, also
double-lobed radio source giving rise to a so-called double-double radio
galaxy -- DDRG \citep{lara99, schoen00a, schoen00b}. According to the theory
developed by \citet{kaiser00} -- hereafter KSR -- one of the most noticeable
features of DDRGs is that they are large-size radio objects ($\ga 700$ kpc)
because recurrent activity cannot be observed in radio structures that are
significantly
smaller than GRGs. The reason for this is that $\sim$$5\times10^7$ years are
required to refill the cocoon with matter from warm clouds of gas embedded in
the hot Intergalactic Medium (IGM) passing through the bow shock surrounding
the outer source structure. Otherwise, the gas densities within the cocoon are
insufficient for the development of new shocks/hotspots.

In spite of this, an attempt was made to search for signatures of
restarted activity in radio sources with sizes less than those of GRGs. This is
important because, if DDRGs with linear sizes considerably below 1\,Mpc {\em
do} exist, modification of the theory of KSR would appear inevitable. In
Sect.~\ref{s-search} a procedure is presented to search for candidates that are
potentially restarted objects with sizes below 1\,Mpc. Using this
procedure a small sample of such objects was constructed and observed with
MERLIN at 5\,GHz (Sect.~\ref{s-obs}). The results are
discussed in Sect.~\ref{s-disc} and summarised in Sect.~\ref{s-concl}.


\section{A search for restarted objects in the FIRST survey}\label{s-search}

Radio sources with extended emission, whose structures have been clearly imaged
in the {\em Faint Images of Radio Sky at Twenty centimetres} (FIRST) survey
\citep{wbhg97}\footnote{Official website: http://sundog.stsci.edu}, have
angular sizes of the order of a few arcminutes and therefore linear sizes less
than $\approx$1\,Mpc, regardless of their redshifts. Consequently, an
initial systematic search through the FIRST catalogue for triple sources with a
dominant core would yield objects with sizes less than 1\,Mpc, which could form
a list of candidates for objects showing signatures of restarted activity.

An automatic procedure was developed for this. In the first phase of this
procedure, a source within FIRST with a flux density greater than a particular
limit (75\,mJy) was selected and a pair of ``secondary'' sources (the ``lobes'')
 were sought within a $2\arcmin$ radius of the initially selected source (the
``core''). An aligned triple was assumed to have been detected if the angle
lobe$_1$~--~core~--~lobe$_2$ was greater than $165\degr$ and the peak flux
densities of the ``lobes'' were less than 30\% of that of the ``core''. This
was repeated for all the sources in FIRST with flux densities that were greater
than the limit.
The 30\% threshold was somewhat arbitrary, but was based upon an initial,
extensive visual inspection of FIRST maps: the peak flux densities of the lobes
in the vast majority of sources with the required morphology fell well below
that limit anyway. No lower limit for the flux density of the FIRST components
of relic lobes was imposed. The sources of this kind are labelled
core-dominated triples (CDTs) throughout this paper.

As a next step, 5-GHz flux densities of the selected sources were obtained from
the GB6 survey \citep{GB6}, and those sources with flux densities below 40\,mJy
were rejected. Also rejected were those whose spectral indices calculated from
the FIRST and GB6 flux densities were: $\alpha < 0.5$
($S\propto\nu^{-\alpha}$). The 5-GHz flux density limit was adopted to ensure
that high signal to noise images of the ``cores'' would result from MERLIN
follow-up observations at that frequency. (The aforementioned 75\,mJy limit at
1.4\,GHz resulted from the 5-GHz flux density limit and the spectral index
criterion.) The sample was limited to objects with steep spectra to maximise
the probability that the ``cores'' themselves, which mainly contribute to the
total flux density, were likely to have steep spectra\footnote{Obviously,
calculation of the
spectral indices of the ``cores'' {\em alone} was not possible because of the
resolution limit of the GB6 survey.}. Generic core-jet sources, which typically
have flat spectra, were possibly eliminated in this way.

This automated procedure was carried out using the version of the FIRST
catalogue dated 15~October 2001 (the most recent at the time the project began)
covering approximately 7954 square degrees in the north Galactic cap and 611
square degrees in the south Galactic cap, and containing $\sim$771,000 sources.
It follows that with the GB6 covering only the northern hemisphere, our
sample is also limited in the same manner.
It yielded 313~sources; their maps were extracted from the FIRST archive and
inspected visually. It turned out that the majority of the alleged triple
structures were typically compact objects surrounded by symmetric artefacts
originating from the VLA diffraction pattern that is typical of snapshot mode
observations. These were rejected and a final sample of 15~sources resulted
from the whole selection process. Their names, coordinates, flux densities at
1.4\,GHz and 4.85\,GHz, and spectral indices between these two are shown in
Table~\ref{t-sample}.

The approach described above is somewhat complementary
to the one of \citet{schoen00c} and \citet{lara01}. These authors searched for
GRGs in WENSS and NVSS, respectively, and their selection criteria included a
lower limit to the sources' angular sizes, which was $5\arcmin$ and $4\arcmin$,
respectively. As mentioned in Sect.~\ref{s-intro}, in this study,
objects that are less extended than GRGs are being investigated and so an
upper limit of $4\arcmin$ has been adopted.

After MERLIN observations of the 15 targets (reported in Sect.~\ref{s-obs})
had been made, four sources of greatest interest were chosen for more detailed
analysis. In Table~2 
the flux densities and locations of
the FIRST components of these four are quoted from the FIRST catalogue.

\begin{figure*}[t]
\includegraphics[scale=0.4]{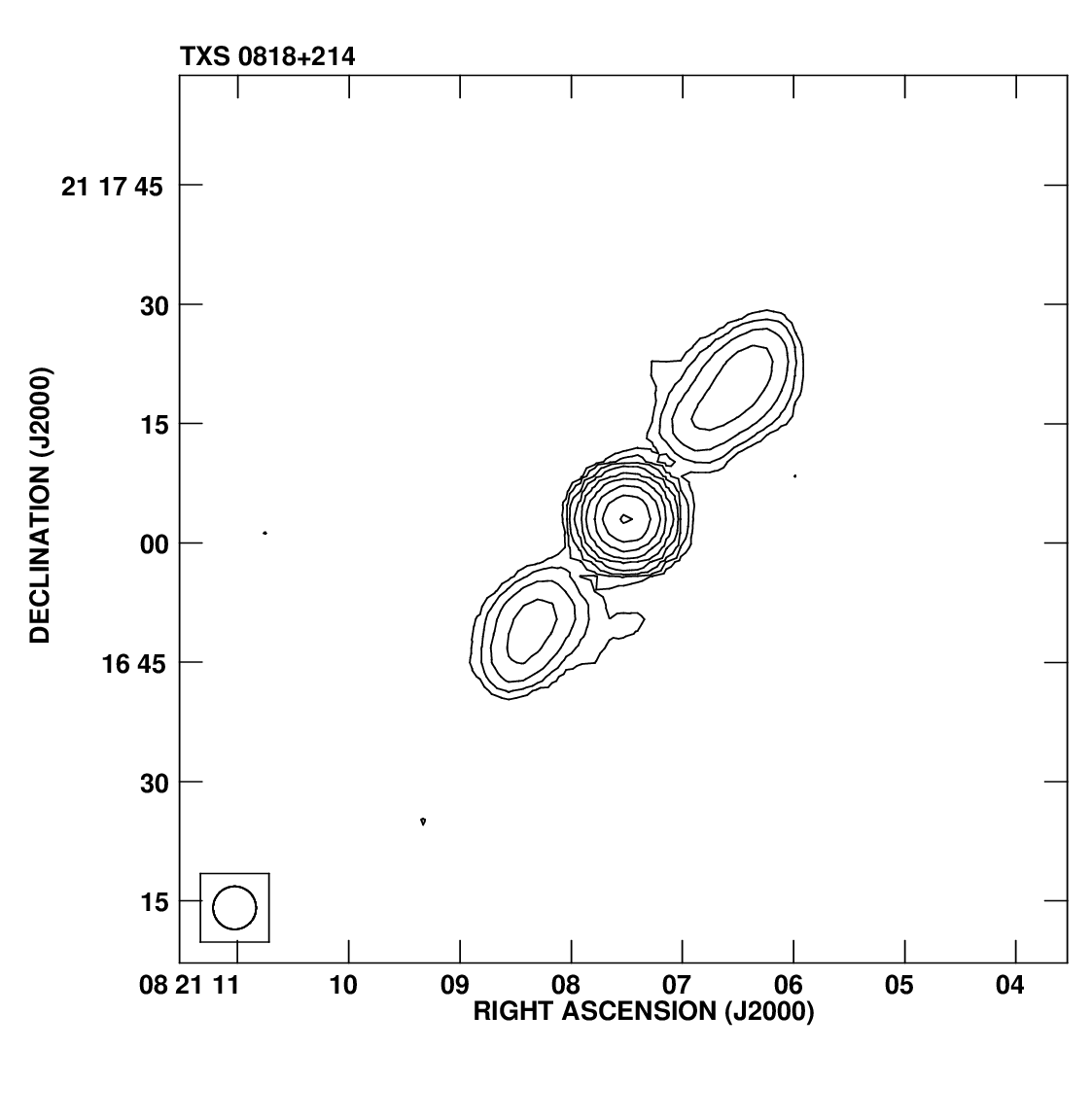}
\includegraphics[scale=0.4]{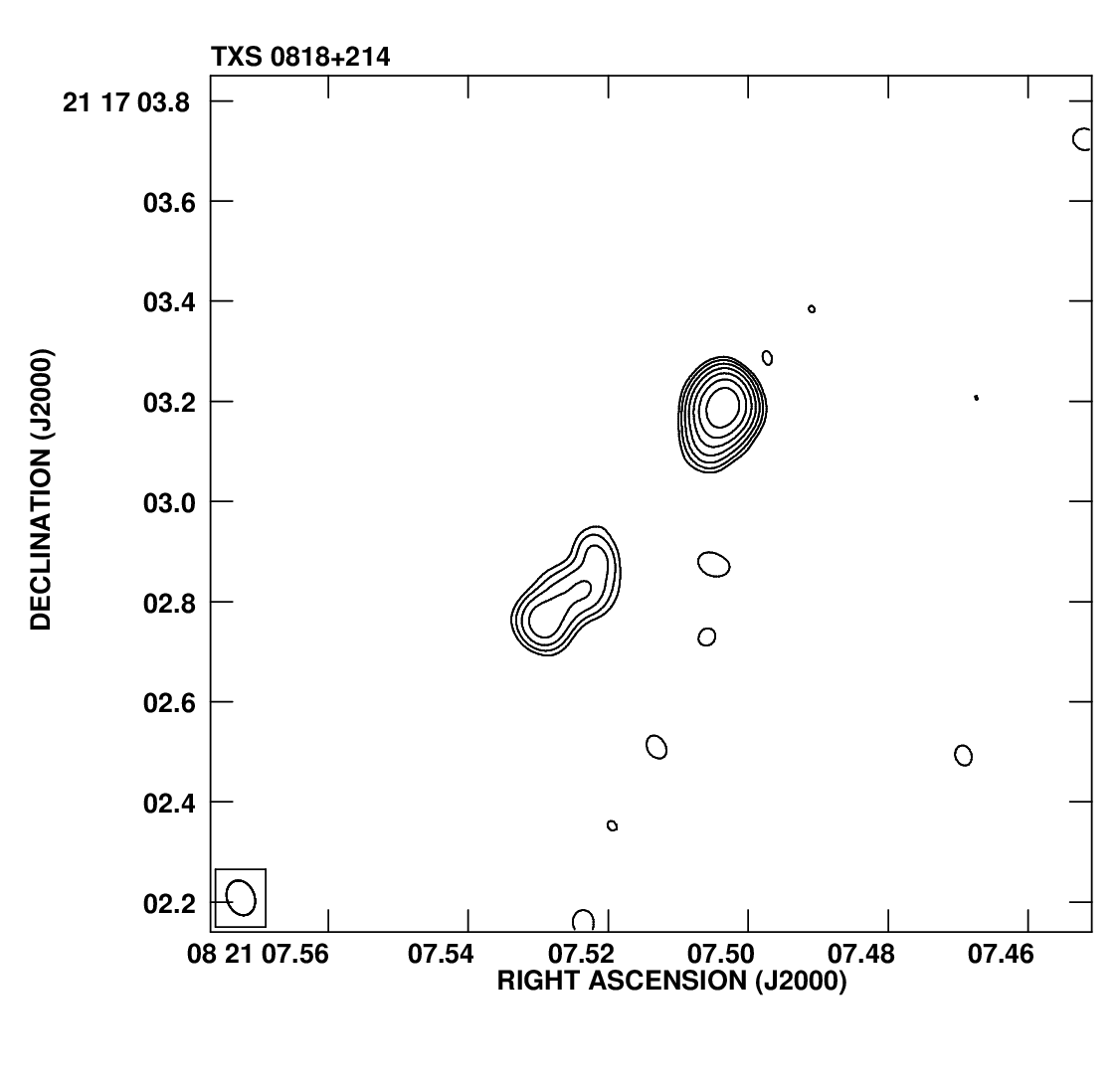}
\caption{TXS\,0818+214. {\em Left panel}: FIRST map. Contours increase by
a factor of 2; the first contour level is at 0.5\,mJy/beam.
{\em Right panel}: MERLIN map at 5\,GHz. Contours increase by a
factor of 2; the first contour level corresponds to $\approx 5\sigma$ level,
which is 0.23\,mJy/beam. The peak flux density is 27.3\,mJy/beam.
The beam size is $72 \times 54$\,milliarcseconds at a position
angle of $21\degr$.}
\label{fig:0818+214}
\end{figure*}

\begin{table}[t]
\caption[]{Fitted parameters of the MERLIN map components.}
\begin{flushleft}
\begin{tabular}{c c c r}
\hline
\hline
Source &      &     & ${\rm S_{5GHz}}$\\
name   & R.A. & Dec & [mJy]\\
(1)    & (2) & (3)   & (4)\\
\hline
TXS\,0818+214 & 08 21 07.504 & 21 17 03.18 &  39.9\\
              & 08 21 07.526 & 21 17 02.80 &   9.0\\
TXS\,1033+026 & 10 36 31.946 & 02 21 45.54 &  11.4\\
              & 10 36 31.938 & 02 21 45.41 &   7.5\\
TXS\,1308+011 & 13 11 21.157 & 00 53 19.47 &  15.2\\
              & 13 11 21.161 & 00 53 19.26 &  39.3\\
TXS\,1312+563 & 13 14 58.405 & 56 03 42.43 &  78.8\\
              & 13 14 58.412 & 56 03 41.81 &  46.6\\
\hline                                                
\end{tabular}
\end{flushleft}

{\small
Description of the columns:
Col.~(1): Source name;
Col.~(2): Component right ascension (J2000) estimated using JMFIT;
Col.~(3): Component declination (J2000) estimated using JMFIT;
Col.~(4): Flux density (mJy) obtained using JMFIT;
}

\label{MERLIN_JMFIT}
\end{table}


\section{Observations}\label{s-obs}

\subsection{Overview}

The MERLIN full-track, phase-referenced observations at 5\,GHz were carried out
for the sample of 15~sources during the period 16 Feb. -- 1 Jun. 2004. 3C286 was
used as the flux density and polarisation calibrator, and OQ208 was used as the
point source calibrator. The flux density of 3C286 on the VLA scale, which was
used, was 7.331\,Jy. The initial editing and calibration of the data were
carried out using the Jodrell Bank d-programmes, and preliminary
phase-referenced images were produced using the AIPS based PIPELINE procedure
developed at Jodrell Bank. Further cycles of phase self-calibration and imaging
using the AIPS tasks CALIB and IMAGR were then used to produce the final images.

As indicated in Table~\ref{t-sample}, two of the five weakest sources of
the sample with GB6 flux densities below 50\,mJy -- TXS\,1024+549 and
TXS\,1025+089 -- were not detected at the $3\sigma$ level, corresponding
to 430$\mu$Jy/beam and 320$\mu$Jy/beam, respectively. Another two of these
five -- TXS\,1046+187 and FIRST\,J155726.1+360133 -- were only marginally
detected; i.e. some diffuse structures appear at the $3\sigma$ level
corresponding to 300$\mu$Jy/beam and 390$\mu$Jy/beam, respectively. The lack of
any distinct features in their MERLIN images is probably the result of the
diffuse
structures being resolved. Seven sources appeared to be unresolved or barely
resolved. They are listed as ``single'' in Table~\ref{t-sample} and will not
be discussed further here. The remaining four targets, listed in 
Table~2, 
have well resolved structures consisting of more
than one component. Their MERLIN images are presented and comments on them
given in Sect.~\ref{s-comm}. The flux densities of the components shown in
these images were measured using the AIPS task JMFIT and are listed in
Table~\ref{MERLIN_JMFIT}.

\begin{figure*}[ht]
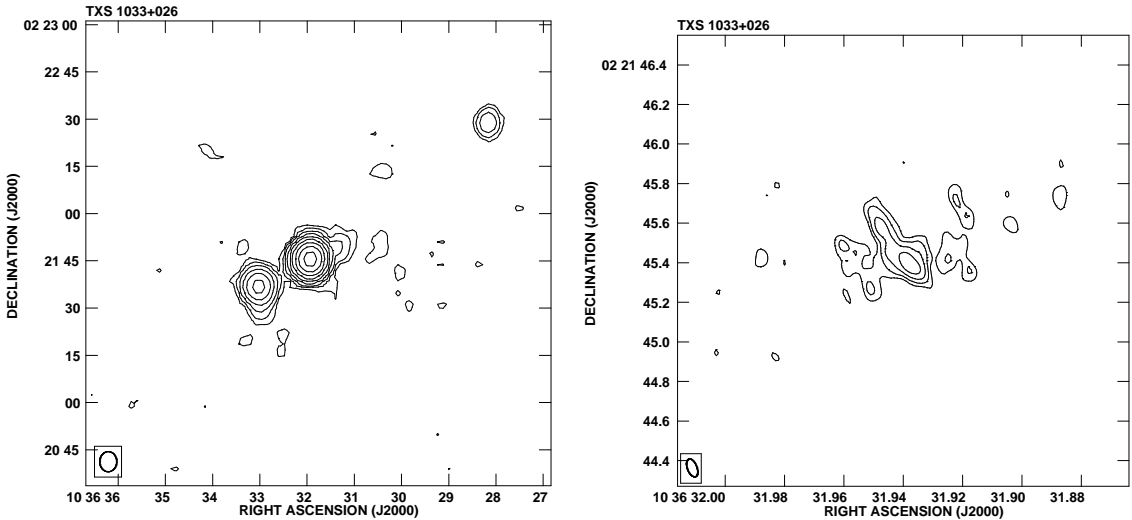

\includegraphics[scale=0.4]{3994fg2a.ps}
\includegraphics[scale=0.4]{3994fg2b.ps}
\caption{TXS\,1033+026. {\em Left panel}: FIRST map. Contours increase by
a factor of 2; the first contour level is at 0.5\,mJy/beam.
{\em Right panel}: MERLIN map at 5\,GHz. Contours increase by a
factor of 2; the first contour level corresponds to $\approx 3\sigma$ level,
which is 0.38\,mJy/beam. The peak flux density is 5.4\,mJy/beam.
The beam size is $96 \times 48$\,milliarcseconds at a position
angle of $21\degr$.}
\label{fig:1033+026}
\end{figure*}

\subsection{Notes on individual sources}\label{s-comm}

\noindent {\bf \object{TXS\,0818+214}}. This object
was included in Release~4 of the Sloan Digital Sky Survey
(SDSS/DR4)\footnote{This is the most recent release at the time of writing.}
as a $m_R=21.73$ galaxy; but its redshift was not known, and it could not be
found in the literature. Based both on the $r$-$z$ Hubble dia\-gram as derived
by e.g. \citet{thom94} and on the given magnitude, an estimated redshift would
be $z\approx1.0$. The MERLIN image of the ``core'' of the source
(Fig.~\ref{fig:0818+214}, right panel) shows it as a double located at
P.A.$=-40\degr$. Given that the two components are asymmetric -- the
northwestern one is appreciably more compact than the southeastern one --
it could be that the subarcsecond-scale structure is a core-jet type.
The approximate alignment between the subarcsecond-scale and arcminute-scale
structures (Fig.~\ref{fig:0818+214}) agrees with this. However,
despite the compactness of the northwestern component, it and the
southeastern part of the source could still be regarded as two FR\,II-type
\citep{fr74} lobes. No measurable polarisation was observed with MERLIN.
Thus, only a detailed study of their spectral indices based on the present
data and future observations of the source at a lower frequency and with a
similar resolution might determine which is the correct interpretation.
However, if the two components are actually lobes and given that the redshift
estimate for this source translates to a linear size $\la 586$\,kpc, such a
linear size would be somewhat less than the minimum for a double-double source
required by the KSR model ($\approx$700\,kpc). The sizes quoted above are not
likely to be underestimated, as 586\,kpc would result from an assumption of
$z=1.25$ and so it is the maximum possible linear size for the observed angular
size ($68\arcsec$) and the adopted cosmological model. Therefore, if the
double-double morphology is confirmed, it would appear that the KSR model
would need to be modified to become applicable to TXS\,0818+214.

\begin{figure}[t]
\centering
\includegraphics[scale=1.0]{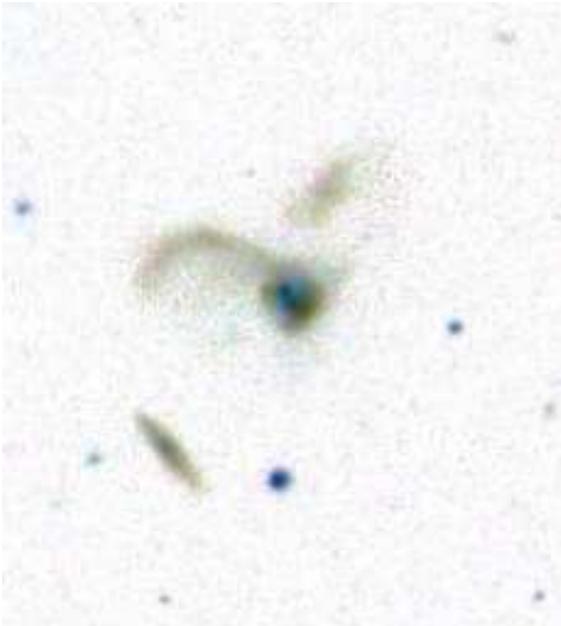}
\caption{Negative of the SDSS/DR4 optical image centred on the position of
the radio source TXS\,1033+026. The size of the image is approximately
$1\arcmin\times 1\arcmin$.}
\label{fig:1033+026.SDSS}
\end{figure}

\noindent {\bf \object{TXS\,1033+026}}. Fig.~\ref{fig:1033+026.SDSS} shows
a $1\arcmin\times 1\arcmin$ (approx.) cutout from the SDSS/DR4 archive centred
on the brightest (``core'') component of TXS\,1033+026 seen in the FIRST image
(Fig.~\ref{fig:1033+026}, left panel).
In the optical domain it is also the brightest feature in this field: a
``fuzzy'' object with two emission peaks in its very centre
and an arc-like structure emerging from there towards the northeast and bent
so that it eventually points to the southeast. There are also two regularly
shaped galaxies in the field: SDSS\,J103631.50+022201.9 with $m_R=17.84$
located $18\arcsec$ to the north of the centre and SDSS\,J103633.24+022120.6
with $m_R=17.63$ located $32\arcsec$ to the southeast of the centre. Neither
of these has a known redshift nor a radio counterpart in the FIRST image.

According to SDSS/DR4, one of the optical peaks, SDSS\,J103631.95+022145.7
(ObjId=587726032776265858), is coincident within 0\farcs4 with the ``core''
of the FIRST source, but its redshift is not listed there, nor could it be
found in the literature. The second optical peak, SDSS\,J103631.87+022144.0
(ObjId=587726032776265859), which is 1\farcs67 off the
position of the radio source, is a galaxy
CGCG\,037$-$089 whose redshift, $z=0.050276$, has been listed in the SDSS
archive. At this redshift an angular distance of $2\arcsec$, which is the
distance between the two optical maxima, corresponds to 2.7\,kpc. At such a
short (projected) distance, one would expect these two objects to be strongly
interacting. Together with the presence of the arc-like structure closely
resembling e.g. those in the ``Antennae'' galaxies, the observed structure
is perhaps evidence that a merger is taking place.

\begin{figure*}[t]
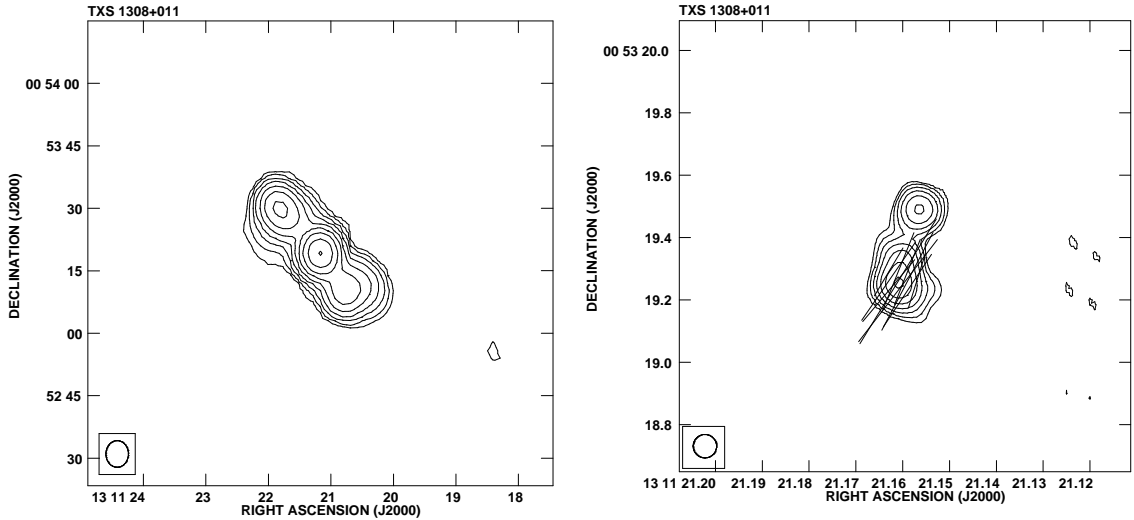

\includegraphics[scale=0.4]{3994fg4a.ps}
\includegraphics[scale=0.4]{3994fg4b.ps}
\caption{TXS\,1308+011. {\em Left panel}: FIRST map. Contours increase by
a factor of 2; the first contour level is at 0.5\,mJy/beam.
{\em Right panel}: MERLIN map at 5\,GHz. Contours increase by a
factor of 2; the first contour level corresponds to $\approx 3\sigma$ level,
which is 0.35\,mJy/beam. The peak flux density is 24.9\,mJy/beam.
The polarisation line of $0\farcs5$ amounts to 2.5\,mJy/beam.
The circular beam size is 75\,milliarcseconds.}
\label{fig:1308+011}
\end{figure*}

\begin{figure*}[t]
\includegraphics[scale=0.4]{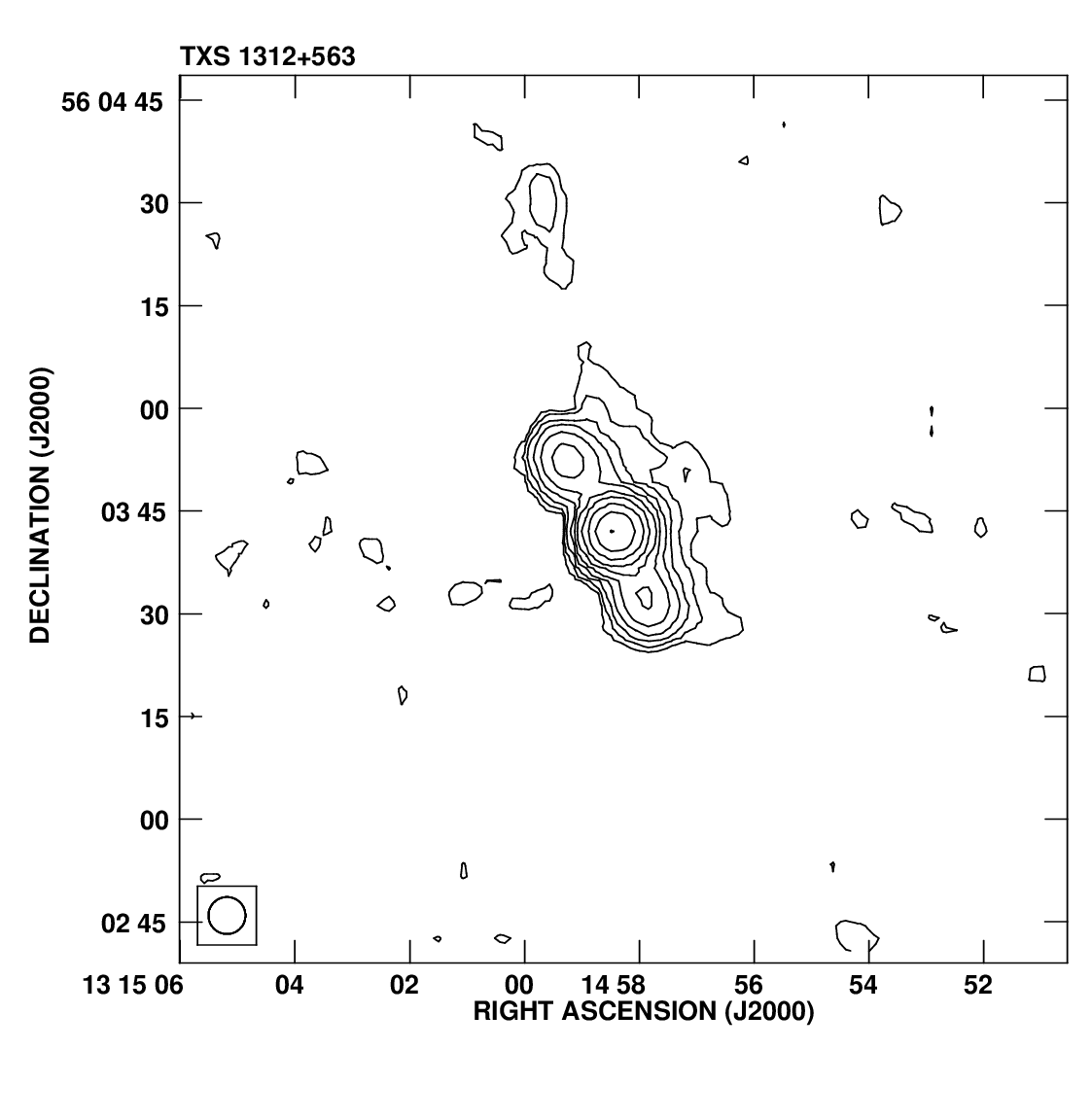}
\includegraphics[scale=0.4]{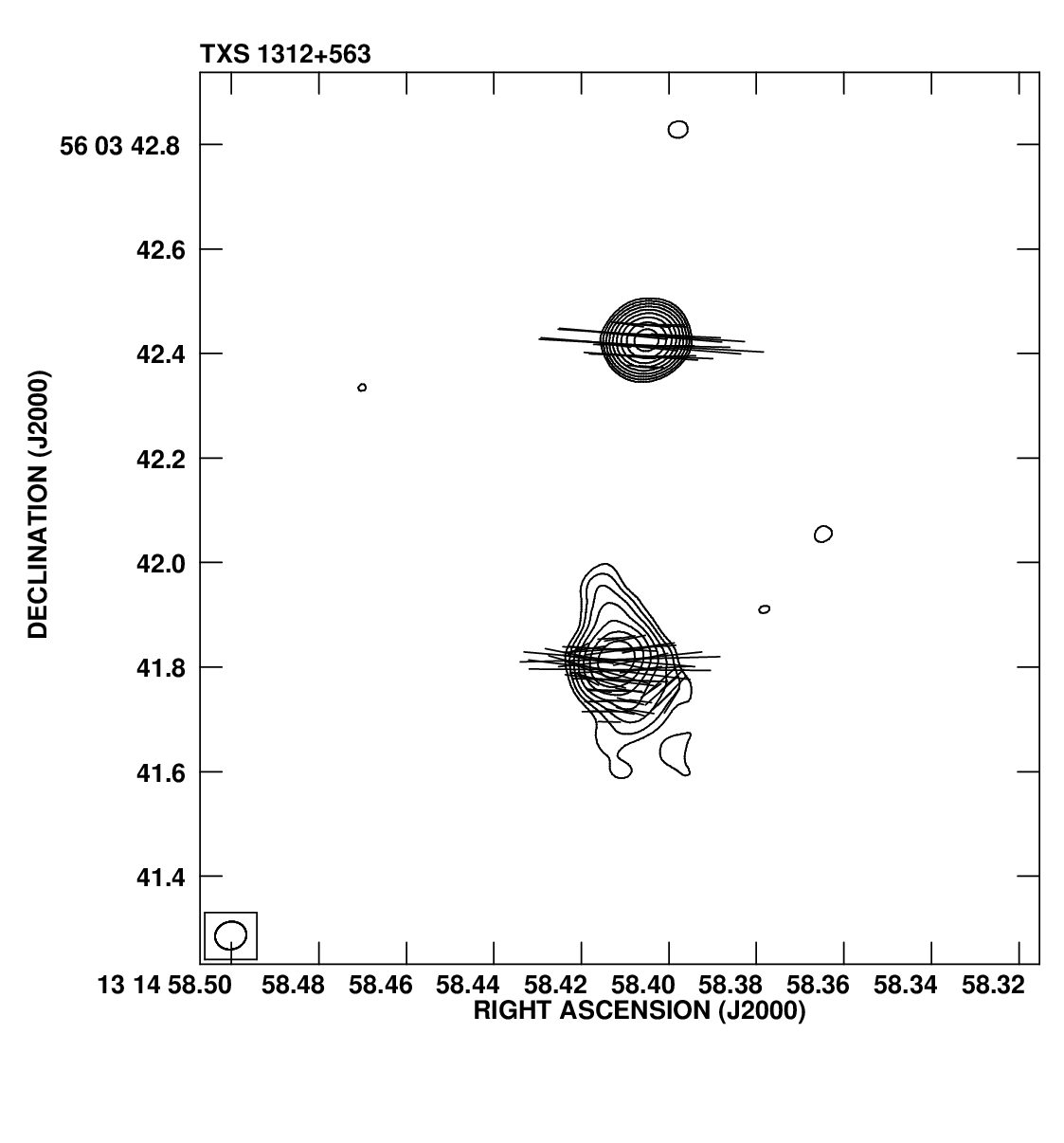}
\caption{TXS\,1312+563. {\em Left panel}: FIRST map. Contours increase by
a factor of 2; the first contour level is at 0.5\,mJy/beam.
{\em Right panel}: MERLIN map at 5\,GHz. Contours increase by a
factor of 2; the first contour level corresponds to $\approx 5\sigma$ level,
which is 0.23\,mJy/beam. The peak flux density is 77.7\,mJy/beam.
Polarisation line of $0\farcs5$ amounts to 2\,mJy/beam.
The beam size is $60 \times 53$\,milliarcseconds at a position
angle of $-76\degr$.}
\label{fig:1312+563}
\end{figure*}

The arcminute-scale structure of the radio source as shown in the FIRST
image (Fig.~\ref{fig:1033+026}, left panel) could be viewed in more than one
way. It could be considered as a very asymmetric triple with a central core, a
lobe to the southeast, and a jet almost pointing in a northwesterly direction
to a second lobe $\simeq$3.5 times farther away from the core than the
southeastern lobe. At the redshift given above, the projected size of the
source would be 141\,kpc. It could also be that the component located at
RA=\,$10^{\rm h}36^{\rm m}28^{\rm s}$, Dec=\,$02\degr 22\arcmin 30\arcsec$
(J2000) is not associated with the main source, but is a coincidence. In
this case, 1033+026 could still be regarded as a CDT with the protrusion
from the core to the northwest being the lobe. The linear size of the
triple would then shrink to 72.6\,kpc.

The principal features of the subarcsecond-scale structure seen in the MERLIN
image (Fig.~\ref{fig:1033+026}, right panel) are a core and a jet pointing to
the northeast. In addition, the ``remains'' of flux to the southeast and to
northwest of the core (i.e. the directions of the two lobes or lobe and jet
visible in the FIRST image), which have been mostly resolved by MERLIN, are
also visible. The jet is at an angle of $\approx 80\degr$ relative to the
direction of the southeastern lobe in the FIRST image.
No measurable polarisation has been observed for this source with MERLIN.

\noindent {\bf \object{TXS\,1308+011}}. This source is identified with a QSO
at a redshift $z=1.075$ \citep{hew95}. The projected linear size is
therefore 420\,kpc. The MERLIN image (Fig.~\ref{fig:1308+011}, right panel)
indicates that in the subarcsecond scale the central component seen in the
FIRST image is a core-jet type source with the northern component, which
appears to be unresolved, being the core. The inner core-jet structure has a
linear size of 3.8\,kpc with the jet initially pointing to the southeast at
P.A.$\approx$$-17\degr$ and then bending away from its $55\degr$ misalignment
relative to the FIRST image structure towards a southwesterly direction, i.e.
into the direction of the lobe in the FIRST image. Significant polarisation
occurs where the jet is bending. There is no indication of a counter-jet to
the north of the core, which is perhaps not visible as it is moving away from
the observer. From the symmetry of the FIRST image, one might expect that any
counter-jet would perhaps bend towards the northeastern component. If all the
morphological features suggested above were actually observed, TXS\,1308+011
could be regarded as an ``Z-shaped'' radio source analogous to TXS\,1346+268
\citep[also known as 4C\,+26.42,][]{bhm84} and a helical jet or precessing jet
model might be applied to it.

The phenomenon of a misalignment between the milli\-arcsecond- and
arcsecond-scale
structures has been thoroughly investigated; see \citet{appl96} for a review.
Based on a large collection of observational material, they have concluded that,
although it is possible to apply a helical jet model to individual sources, it
is difficult to find a single mechanism responsible for the observed
distribution of misalignments, which shows a statistically significant peak
at $\Delta PA\simeq 90\degr$. This peak is particularly well-defined for
BL~Lac objects, but EVN and MERLIN observations of bent radio jets in some
BL~Lac objects \citep{cass02} do not support a helical jet explanation of
misalignments close to $90\degr$. Thus, although an interpretation of the
morphology based on a twisted jet model can be argued -- it has been for
example for TXS\,1055+018 \citep{arw99} -- adopting a scenario of a central
engine repositioning leads to a simple and natural explanation without complex
modelling. Therefore, the restarted activity scenario is not ruled out for
TXS\,1308+011.

\noindent {\bf \object{TXS\,1312+563}}. According to SDSS/DR4, this source is
identified with a QSO at a redshift $z=1.751$. The projected linear size is
therefore 327\,kpc. The MERLIN image (Fig.~\ref{fig:1312+563}, right panel)
of the arcminute-scale ``core'' component shows it as a double source located
at P.A.$\approx$$-5\degr$ with a linear size of 8\,kpc. The northern component
is very compact and is not resolved by MERLIN, but the degree of its
polarisation is appreciable: 2.8\%. The southern component resembles an
FR\,II lobe and is strongly polarised (7.1\%) in the same direction as the
northern one. As in the case of TXS\,0818+214, without spectral index data
it is not possible to classify the inner structure of this source, which could
be either of a core-lobe type as hinted by the morphology or a double-lobe
type as suggested by the polarisation properties.

There is a $30\degr$ misalignment of the subarcsecond-scale double
relative to the arcminute-scale triple structure (Fig.~\ref{fig:1312+563},
left panel). It should be noted that a very similar misalignment has been
observed in the double-double GRG\,1245+676 \citep{mar03}. Thus, a hypothesis
that TXS\,1312+563 could be of a double-double type cannot be ruled out, and if
confirmed, the estimate for the linear size of TXS\,1312+563, which is well
below the $\approx$700\,kpc limit required by the KSR, would indicate
that their theory, unless modified, might not be applicable to this object.


\section{Discussion}\label{s-disc}

As already mentioned in Sect.~\ref{s-intro}, \citet{lara99} and
\citet{schoen00a, schoen00b} have identified a number of radio sources where
the interruption of activity has had a spectacular impact on their
double-double morphologies. J0116$-$47 \citep{sarip02}, PKS\,B1545$-$321
\citep{sarip03}, and SGRS\,J2159$-$7219 \citep{sarip05} are also clear
examples of restarted DDRGs. Three other sources investigated by
\citet{sarip05} -- SGRS J0143$-$5431, J0746$-$5702, and J1946$-$8222 --
show evidence of either one- or two-sided knots/jets closer to the nucleus
and might be examples of giant radio sources with relic lobes and restarting
beams.

Another well-known
type of a dramatic event in the evolu\-tion of a radio-loud AGN that may be
labelled ``restarted activity'' can also lead to the development of an X-shaped
radio source \citep[and references therein]{helgephd}. It is to be noted that
typical X-shaped sources, such as 3C\,223.1 or 3C\,403 \citep{dt02, capetti02},
in principle resemble DDRGs like J0116$-$473 or PKS\,B1545$-$321 except for the
misalignment between the active (inner) and the inactive (outer) parts.
Thus, it might be speculated that the mechanism that triggered the development
of the new structures in these two kinds of objects is the same, but that
DDRGs are just ``special cases'' where the misalignment is (close to) zero.
However, the opposite suggestion, namely that different mechanisms could
be responsible for activity renewal in each type of object, appears to
be more justified at the present time. A review of several possible mechanisms
has been given by \citet{schoen00a}.

Mergers seem to be the most ``natural'' explanation of activity re-ignition,
but if the angular momentum of the infalling gas is different from that of
the pre-existing accretion disk, which is what would normally be
expected\footnote{Consequently, well-aligned DDRGs should be exceptional.
How\-ever, \citet{liu03} point out that if the mass ratio of the two merging
black holes is in the range 0.01 to 0.4, realignment of the central engine
and the subsequent formation of an X-shaped radio structure are not inevitable,
and so the above-mentioned speculation that DDRGs can be regarded as 
``zero-misalignment'' X-shaped sources could still be a viable scenario.},
a fast realignment of the jet axis might occur and an X-shaped source could
emerge (KSR). \citet{me02} show that the orientation of a black hole spin axis
would change dramatically even in a minor merger, leading to an almost
instantaneous reorientation of the coalescing SMBHs and a sudden flip in the
direction of the jets. \citet{liu04} suggests that the realignment of a
rotating SMBH with a misaligned accretion disk is due to the Bardeen-Petterson
effect \citep{bp75} and that the timescale of such a realignment is
$<10^5$\,years. It follows that the change of the direction of the jets is
abrupt compared with the typical age of the radio sources \citep{kg94}.

However, it is to be noted, that, although mergers seem to be the simplest
explanation of activity re-ignition, an alternative model based upon the 
existence of thermal-viscous instabilities in the accretion disks of AGNs 
has been proposed \citep[][and references therein]{hse01}. According to
\citet{kb05}, \citet{mar06}, and \citet{kb06},
this model could explain the existence of sources in which activity has ceased,
but this does not preclude the possibility that the same mechanism could
be responsible not only for a cessation but also for a re-ignition of activity.

The most outstanding feature of both DDRGs and X-shaped sources is that they
are large -- their linear sizes exceed 700\,kpc -- which indicates that they
{\em have} to be particularly old for activity renewal regardless of their
morphological features. The apparent lack of observable double-double 
structures in small-scale sources could mean that specific conditions are
required for activity re-ignition, which do exist in GRGs, but which make it
rare if not impossible for such a phenomenon to occur in more compact
objects. It could be that the physical conditions inside the cocoon might
not favour the development of the inner lobes for a long time after the
initial burst of the activity. According to KSR, that timescale could be up to
$\sim$$5 \times 10^7$ years so that, even if the activity in a source
younger than that {\em could} actually cease and than restart, such an
event would remain unobservable.

While the above currently appears to be the case for DDRGs without exception,
X-shaped sources with linear sizes significantly below 1\,Mpc may exist.
Two examples of such sources -- TXS\,0229+131 \citep{murphy93} and 4C\,+01.30
\citep{wang03} -- can be found in the literature. The arms of TXS\,0229+131
($z=2.059$) have lengths in the range 100 -- 120\,kpc. (The strange appearance
of TXS\,0229+131 could in principle result from gravitational lensing but
according to Browne, priv. comm., it is very unlikely.) The ``active'' and
``relic'' arms of 4C\,+01.30 ($z=0.132$) are 160 and 300\,kpc in length,
respectively.

The conspicuous X-shaped structures observed in some Mpc-scale radio sources
and the two more compact ones mentioned above require a lack of almost any
distortions caused by orientation effects. In other words, both arms of the
cross must lie close to the sky plane for sources to be observed as such. If
such an exceptional orientation is not present, Doppler boosting and beaming
effects would distort the X-shaped structure considerably. In particular, if
the arms pertinent to the ``old'' structures lie in the sky plane whereas
the arms resulting from the renewed activity do not, the observed structure
would be a superposition of an ``old'', double-lobed fader-like component
and an asymmetric i.e. core-jet central component, which might not be perceived
as such in low-resolution observations that are sufficient for a proper imaging
of the double-lobed component. It follows that CDTs could actually be X-shaped
sources distorted by beaming and Doppler boosting, and their morphologies are
a consequence of a fast repositioning of the central engine. Recently, as a
result of investigating ZwCl\,0735.7+7421, \citet{cohen05} have
suggested that a realignment of a rotating SMBH followed by a repositioning of
the accretion disk and the jets is a plausible interpretation of misaligned
radio structures, even if they are {\em not} conspicuously X-shaped.

We conclude that it is plausible that three misaligned radio structures shown
here in detail could be distorted X-shaped or Z-shaped sources probably
resulting from merger events. The existence of a merger in TXS\,1033+026
actually seems to be visible in the SDSS image. It is to be noted that the
theory given by KSR, which precludes the possibility of the existence of
aligned mini-DDRGs, can be reconciled with observations of misaligned restarted
sources with outer structure sizes below $\approx$700\,kpc. This is because the
misalignments are usually large and the shapes of the cocoons highly
non-spherical. In such circumstances, even if the parameters of the medium
inside the cocoon prevent the development of inner lobes so that the secondary
jets move ``silently'' across the cocoon's interior, they would shortly reach
the side of the cocoon where they would be exposed to the unperturbed and much
higher density IGM. Observable jets or hotspots could then appear.


\section{Conclusions and future work}\label{s-concl}

A programme to find radio sources with linear sizes below 1\,Mpc and showing
signatures of possibly restarted activity was undertaken. To this end a
computer-aided selection procedure aimed at finding CDTs in the FIRST catalogue
was devised and a sample of candidates drawn up. These were then observed with
MERLIN at 5\,GHz to reveal the subarcecond-scale morphology of their central
(``core'') components. In 4 out of 15 targets observed, the ``cores'' appear
either as doubles or core-jets, with three of them misaligned with respect to
their outer structures. Repositioning of the central engine of an AGN caused
by a merger event seems to be a plausible scenario leading to the development
of the observed morphologies in these three sources, as difficulties
can arise with application of the helical jet model when attempting to explain
large misalignments \citep[and references therein]{cass02}. Under such
circumstances the concept of restarted activity, coupled with a swift
repositioning of the central engine of the radio source due to a merger event,
is a very competitive alternative because of its simplicity and plausibility.
As this scenario may also work very well for X-shaped sources, they could
be regarded as a parent population of CDTs; the apparent existence of the
latter class is simply the result of orientation effects on X-shaped sources.
More numerous and complete samples of both classes are needed to confirm this
on a sound statistical basis but, at least, the scarcity of both X-shaped and
CDT objects seems to qualitatively agree. 

No misalignment was detected in the fourth source under investigation,
TXS\,0818+214, and only a rather modest ($30\degr$) misalignment was observed
in TXS\,1312+563. The inner parts of these two sources could be interpreted
as double-lobed structures, but further multifrequency observations are
necessary to provide support for such an interpretation. If confirmed,
TXS\,0818+214 and TXS\,1312+563 would be the most compact double-double
radio sources known to date. An attempt to explain their nature by the theory
developed by KSR would make a modification of this theory necessary, given that
the linear sizes of these two sources are below the required limit.

\begin{acknowledgements}

\item MERLIN is operated by the University of Manchester as a National Facility
on behalf of the Particle Physics \& Astronomy Research Council (PPARC).

\item This research has made use of the
NASA/IPAC Extragalactic Database (NED) which is operated by the Jet
Propulsion Laboratory, California Institute of Technology, under contract
with the National Aeronautics and Space Administration.

\item Use has been made of the fourth release of the Sloan Digital Sky Survey
(SDSS) Archive. Funding for the creation and distribution of the SDSS
Archive has been provided by the Alfred P. Sloan Foundation, the
Participating Institutions, the National Aeronautics and Space
Administration, the National Science Foundation, the U.S. Department of
Energy, the Japanese Monbukagakusho, and the Max Planck Society. The SDSS
Web site is http://www.sdss.org/. The SDSS is managed by the Astrophysical
Research Consortium (ARC) for the Participating Institutions. The
Participating Institu\-tions are The University of Chicago, Fermilab, the
Institute for Advanced Study, the Japan Participation Group, The Johns
Hopkins University, Los Alamos National Laboratory, the
Max-Planck-Institute for Astronomy (MPIA), the Max-Planck-Institute for
Astrophysics (MPA), New Mexico State University, University of Pittsburgh,
Princeton University, the United States Naval Observatory, and the
University of Washington.

\item AM acknowledges the receipt of a travel grant funded by Radio\-Net
as a part of the Trans-National Access (TNA) programmes.

\item We thank Tom Muxlow for help with data reduction.

\end{acknowledgements}

\end{document}